\input harvmac

\noblackbox

\Title{\vbox{\baselineskip12pt\hbox{hep-th/0407130}
\hbox{SU-ITP-04/26, SLAC-PUB-10502, TIFR/TH/04-17} }}
 {\vbox{ {\centerline{Gaugino Condensation and Nonperturbative}
\smallskip
\centerline{Superpotentials in Flux Compactifications }} } }

\centerline{Lars G\"orlich$^a$,
Shamit Kachru$^b$,
Prasanta K.~Tripathy$^a$ and Sandip P.~Trivedi$^a${\foot{emails:
goerlich, prasanta, sandip @theory.tifr.res.in, skachru@stanford.edu}}}
\smallskip
\centerline{$^{a}$ Tata Institute for Fundamental Research}
\centerline{Homi Bhabha Road, Mumbai 400 005, INDIA}
\smallskip
\centerline{$^{b}$ Department of Physics and SLAC}
\centerline{Stanford University}
\centerline{Stanford, CA 94305/94309, USA}

\medskip

\noindent
There are two known sources of nonperturbative superpotentials for
K\"ahler moduli in type IIB orientifolds,
or F-theory compactifications
on Calabi-Yau fourfolds, with flux: Euclidean brane instantons and low-energy
dynamics in D7 brane gauge theories.  The first class
of effects, Euclidean D3 branes
which lift in M-theory to M5 branes wrapping divisors
of arithmetic
genus 1 in the fourfold,
is relatively well understood.  The second
class has been less explored.  In this paper, we consider the explicit
example of F-theory on $K3 \times K3$ with flux.
The fluxes lift the D7
brane matter fields, and stabilize stacks of D7 branes at loci of
enhanced gauge symmetry.
The resulting theories exhibit
gaugino condensation,
and generate a nonperturbative superpotential for K\"ahler moduli. We
describe how the relevant geometries in general contain cycles of
arithmetic genus $\chi \geq 1$ (and how $\chi > 1$ divisors can
contribute to the superpotential, in the presence of flux).
This second class of effects is likely to be important in finding
even larger classes of models where the KKLT mechanism of moduli
stabilization can be realized.  We also address various
claims about the situation 
for IIB models with a single K\"ahler modulus.

\Date{July 2004}

\def\eto{{\hat{\tilde e}_1}}
\def\ett{{\hat{\tilde e}_2}}

\lref\DeAlwis{R. Brustein and S. de Alwis, ``Moduli potentials
in string compactifications with fluxes: Mapping the discretuum,''
hep-th/0402088.}

\lref\DDFK{F. Denef, M. Douglas, B. Florea and S. Kachru, to appear.}
\lref\Diaconescu{D. Diaconescu and S. Gukov, ``Three Dimensional
$N=2$ Gauge Theories and Degenerations of Calabi-Yau Fourfolds,''
Nucl. Phys. {\bf B535} (1998) 171, hep-th/9804059.}
\lref\CascalesQP{
J.~F.~G.~Cascales and A.~M.~Uranga,
``Branes on generalized calibrated submanifolds,'' hep-th/0407132.
}
\lref\CamaraJJ{
P.~G.~Camara, L.~E.~Ibanez and A.~M.~Uranga,
``Flux-induced SUSY-breaking soft terms on D7-D3 brane systems,''
 hep-th/0408036.\semi
P. Camara, L. Ibanez and A. Uranga, ``Flux-induced SUSY-breaking
soft terms,'' Nucl. Phys. {\bf B689} (2004) 195, hep-th/0311241.
}
\lref\Moore{G. Moore, ``Les Houches Lectures on Strings and
Arithmetic,'' hep-th/0401049.}
\lref\BP{R. Bousso and J. Polchinski, ``Quantization of Four-Form Fluxes
and Dynamical Neutralization of the Cosmological Constant,''
JHEP {\bf 0006} (2000) 006, hep-th/0004134.}
\lref\MSS{A. Maloney, E. Silverstein and A. Strominger,
``De Sitter Space in Noncritical String Theory,'' hep-th/0205316.}
\lref\KKLT{S. Kachru, R. Kallosh, A. Linde and S. P.  Trivedi, ``De Sitter
Vacua in String Theory,'' Phys. Rev. {\bf D68} (2003) 046005,
hep-th/0301240.}
\lref\Susskind{L. Susskind, ``The Anthropic Landscape of String
Theory,'' hep-th/0302219.}
\lref\DDF{F. Denef, M. Douglas and B. Florea, ``Building a Better
Racetrack,'' JHEP {\bf 0406} (2004) 034, hep-th/0404257.}
\lref\Douglas{M. Douglas, ``The Statistics of String/M theory vacua,''
JHEP {\bf 0305} (2003) 046, hep-th/0303194.}
\lref\Statistics{S. Ashok and M. Douglas, ``Counting Flux Vacua,''
JHEP {\bf 0401} (2004) 060, hep-th/0307049.}
\lref\StatII{F. Denef and M. Douglas, ``Distributions
of Flux Vacua,'' JHEP {\bf 0405} (2004) 072, hep-th/0404116.}
\lref\GKT{A. Giryavets, S. Kachru and
P. Tripathy, ``On the Taxonomy of Flux Vacua,'' hep-th/0404243.}
\lref\unpub{S. Kachru, unpublished.}
\lref\Sethi{D. Robbins and S. Sethi, ``A Barren Landscape,''
hep-th/0405011.}
\lref\Katz{S. Katz and C. Vafa, ``Geometric Engineering
of ${\cal N}=1$ Quantum Field Theories,''
Nucl. Phys. {\bf B497} (1997) 196, hep-th/9611090.}
\lref\Aspin{ P. S. Aspinwall, ``$K3$ Surfaces and String Duality,''
hep-th/9611137.}
\lref\TT{P. K. Tripathy and S. P. Trivedi, ``Compactifications with Flux
on $K3$ and Tori,'' JHEP {\bf 0303} (2003) 028, hep-th/ 0301139.}
\lref\GKP{S. Giddings, S. Kachru and J. Polchinski,
``Hierarchies From Fluxes in String Compactifications,''
Phys. Rev. {\bf D66} (2002) 106006, hep-th/0105097.}
\lref\Polchinski{J. Polchinski, {\it String Theory, Volume II},
Cambridge University Press, 1998.}
\lref\DRS{K. Dasgupta,
G. Rajesh and S. Sethi, ``M Theory, Orientifolds and $G$-Flux,''
JHEP {\bf 9908} (1999) 023, hep-th/9908088.}
\lref\tadpole{K. Becker and M. Becker, ``M-theory on Eight Manifolds,''
Nucl. Phys. {\bf B477} (1996) 155, hep-th/9605053\semi
S. Sethi, C. Vafa and E. Witten, ``Constraints on low-dimensional string
compactifications,'' Nucl. Phys. {\bf B480} (1996) 213, hep-th/9606122.}
\lref\KS{I. Klebanov and M. Strassler, ``Supergravity and a confining
gauge theory: Duality cascades and $\chi$SB resolution of naked
singularities,'' JHEP {\bf 0008} (2000) 052, hep-th/0007191.}
\lref\KPV{S. Kachru, J. Pearson and H. Verlinde, ``Brane/Flux
Annihilation and the String Dual of a Non-Supersymmetric Field Theory,''
JHEP {\bf 0206} (2002) 021, hep-th/0112197.}
\lref\Witten{E. Witten, ``Nonperturbative
superpotentials in string theory,'' Nucl. Phys. {\bf B474}
(1996) 343, hep-th/9604030.}
\lref\GVW{S. Gukov, C. Vafa and E. Witten, ``CFTs from Calabi-Yau
Fourfolds,'' Nucl. Phys. {\bf B584} (2000) 69, hep-th/9906070;
T. Taylor and C. Vafa, ``RR Flux on Calabi-Yau and Partial Supersymmetry
Breaking,'' Phys. Lett. {\bf B474} (2000) 130, hep-th/9912152;
P. Mayr, ``On Supersymmetry Breaking in String Theory and its
Realization in Brane Worlds,'' Nucl.
Phys. {\bf B593} (2001) 99, hep-th/0003198.}
\lref\Burgess{C.P. Burgess, R. Kallosh and F. Quevedo, ``de Sitter
string vacua from supersymmetric D-terms,'' JHEP {\bf 0310} (2003)
056, hep-th/0309187.}
\lref\Saltman{A. Saltman and E. Silverstein, ``The scaling of
the no-scale potential and de Sitter model building,'' hep-th/0402135.}
\lref\Grassi{A. Grassi, ``Divisors on elliptic Calabi-Yau fourfolds and
the superpotential in F-theory I,'' J. Geom. Phys. {\bf 28} (1998) 289.}
\lref\reviews{E. Silverstein, ``TASI/PITP/ISS Lectures on Moduli and
Microphysics,'' hep-th/0405068;
V. Balasubramanian, ``Accelerating Universes and String Theory,''
hep-th/0404075\semi
A. Frey, ``Warped Strings: Self-Dual Flux and Contemporary
Compactifications,'' hep-th/0308156.}
\lref\meurice{D. Amati, K. Konishi, Y. Meurice and
G. Veneziano, ``Nonperturbative Aspects in Supersymmetric
Gauge Theories,'' Phys. Rept. {\bf 162} (1988) 169.}
\lref\Klemm{A. Klemm, B. Lian, S.S. Roan and S.T. Yau, ``Calabi-Yau
Fourfolds for M-theory and F-theory Compactifications,''
Nucl. Phys. {\bf B518} (1998) 515, hep-th/9701023.}
\lref\Ferrara{C.~Angelantonj, R.~D'Auria, S.~Ferrara and M.~Trigiante,
 ``K3 x T$^2$/Z$_2$ orientifolds with fluxes, open string moduli and critical
points,''
Phys. Lett. {\bf B583} (2004) 331, hep-th/0312019.}
\lref\LustFI{
D.~L\"ust, S.~Reffert and S.~Stieberger,
``Flux-induced soft supersymmetry breaking in chiral type IIb orientifolds with D3/D7-branes,'' hep-th/0406092.}

\newsec{Introduction and Review}

For many years, it has been clear that string theory offers a plethora
of choices for compactification to 4d with ${\cal N}\geq 1$
supersymmetry. With ${\cal N}=1$ supersymmetry, quantum effects can play
an important role in breaking supersymmetry and/or
changing the vacuum structure.  It has long been hoped that such
effects, perhaps in conjunction with early universe dynamics, would
yield one or a few models as the preferred string compactifications --
i.e., that there would be a simple vacuum selection principle.  There is
little evidence for such a picture (though of course our understanding
of early universe cosmology in string theory is very limited).
Instead, recent attempts to understand moduli stabilization have yielded
mounting evidence that after including perturbative and
nonperturbative contributions to the moduli potential,
string theory manifests a tremendous landscape
of vacua,
including 4d (metastable) de Sitter and anti-de Sitter geometries with a
wide range of different cosmological constants
\refs{\BP \MSS \KKLT \Susskind{--} \Douglas}.
For nice reviews of this subject, see \reviews.
While a similar picture should emerge in each of the corners of the
M-theory parameter space, we will concern ourselves here with the type
IIB theory, where the story is best developed.  Because our goal in this
paper is to clarify and extend some recent developments in this area, we
will begin with a short review and status report on the subject.

If one wishes to obtain theories with 4d ${\cal N} \leq 1$ supersymmetry
in the IIB setting, one large class of constructions was developed in
\GKP\ and references therein.  These models are Calabi-Yau orientifolds
with D3 and/or D7 branes, and also admit a description as F-theory
compactified on a Calabi-Yau fourfold.  In a given such construction
corresponding to compactification on the threefold $M$ (with related
fourfold $X_4$), one finds a tadpole condition which requires that the
total D3
brane charge on $M$ add up to zero \tadpole

\eqn\tc{N_{D3} + {1\over (2\pi)^4 (\alpha')^2}
\int_{M} H \wedge F ~=~{\chi(X_4)\over 24} ~.}
Here $H$ and $F$ are the NS and RR three-form field strengths of the IIB
theory, and $N_{D3}$ denotes the number of D3 branes one has chosen to
insert transverse to $M$.
So we see that in generic backgrounds, one will turn on RR and NS
fluxes as part of tadpole cancellation.

The resulting class of 4d ${\cal N}=1$ supergravities was described in
\GKP.  The fluxes generate a superpotential for the complex structure
moduli and the dilaton $\tau$.  Defining $G_3 = F - \tau H$, this
superpotential is of the form \GVW
\eqn\fluxsup{W_{flux} ~=~\int_{M} G_3 \wedge \Omega}
where $\Omega$ is the holomorphic three-form on $M$.
For typical choices of the flux, the dilaton and complex structure
moduli have isolated minima.

On the other hand, the K\"ahler moduli $\rho^i$ of $M$ do not appear in
$W$, and participate in a no-scale cancellation at leading orders in
the $\alpha^\prime$ and $g_s$ expansion \GKP.  One generally
expects that quantum corrections will generate a potential for
the $\rho$ fields.  If this potential arises only from corrections to
the K\"ahler potential, any nontrivial vacua will typically occur for
string scale compactification manifolds, and will be difficult to study.

As emphasized in \KKLT, however, there are at least two classes of
effects that lead one to expect $\rho$-dependent corrections to
the superpotential in many models:

\noindent
1) If $M$ contains a 4-cycle $\Sigma$ with the right topological
properties, Witten argued that Euclidean D3 branes will generate a
nonperturbative superpotential
for the K\"ahler modes controlling the size of $\Sigma$ \Witten.
These cycles lift, in the M-theory fourfold geometry, to
``vertical'' divisors
of arithmetic genus 1 (where ``vertical'' roughly denotes that
they wrap the fiber directions which shrink in the F-theory limit).

\noindent
2) These models typically contain D7 branes.
While as discussed in \DDF\ the D7s often have numerous matter fields in
the most naive Kaluza-Klein analysis of their 4d ${\cal N}=1$
supersymmetric worldvolume gauge theory, the three-form flux can give
masses to many or all of these matter fields.  If the fluxes do this
while stabilizing the D7s at a coincident locus, gaugino condensation
will ensue.  For $N$ D7s wrapping a 4-cycle $\Sigma$, the gauge coupling
will satisfy ${1\over g^2} \sim {\rm Vol}(\Sigma)$, and therefore the
gauge theoretic superpotential will generate a nonperturbative potential
for K\"ahler moduli.

In the approximation that one keeps the leading contribution to this
superpotential, one gets a schematic formula of the form
\eqn\newsup{W = W_{flux} + e^{-a\rho}~.}
If $W_{flux}$ evaluated in the vacuum of the complex structure and
dilaton moduli is ${\it small}$, i.e. $W_{flux} = W_0 << 1$, then one
finds a resulting vacuum for $\rho$ at moderately large volume.  For
instance in \KKLT, a toy example was described that, with $W_0$ of
$10^{-4}$ and $a$ of ${1\over 10}$, achieved
a radius of just above 3 in string units (which translates to $\rho \sim
100$). These moderately large radii can
justify the neglect of more highly damped exponentials in the formula
\newsup, yielding self-consistent solutions to the equations of the
effective field theory.  The existence of solutions with $W_0 << 1$ was
justified in \KKLT\ by explaining that given the number of flux vacua
and naive estimates for how $W_0$ might vary in different solutions,
very small values should arise in a small fraction of the solutions.

The vacua just described are, in the simplest cases, supersymmetric AdS
vacua. It was further argued in \KKLT\ that by e.g. using warped
solutions of the sort described in \GKP\ (which incorporate a
Klebanov-Strassler throat \KS), and including anti-D3 branes (whose
dynamics in such throats was studied in \KPV), one should be able to
obtain de Sitter
solutions to string theory.  Instead of including anti-D3 branes,
one could also imagine using anti-self dual field
strengths in D7 branes \Burgess.
Alternatively, one can simply start in a vacuum of the no-scale
potential which is at positive $V$, and play off the tadpole for
K\"ahler moduli against the nonperturbative corrections \Saltman.
Another possibility was described in \DeAlwis.
The end result, as argued in \KKLT\ and later references, is that the
IIB superstring theory seems to admit a rich landscape of vacua with
many de Sitter and anti-de Sitter critical points, exhibiting different
values of the cosmological constant.  While only a small fraction of the
vacua will arise from small $W_0$ in the language above, and will
therefore arise at moderately large radius, this small fraction was
argued to yield a
large absolute number of vacua.

Two recent lines of development have added considerable support to this
picture. In the first, Ashok and Douglas and later authors
\refs{\Statistics \StatII {--} \GKT} have
studied the statistics
of the flux vacua which arise in the no-scale approximation.  The most
basic result concerns the number of vacua, and finds (as one would
expect from simple generalizations of \BP) that
\eqn\nvac{N_{vac} \sim {1\over b^3(M)!} \left({\chi(X_4)\over
24}\right)^{b^3(M)}.} Typical numbers yield ${\chi \over 24}$ of order
1000 and $b^3(M)$ of order 100, and in fact $N_{vac}$ can be in excess
of $10^{300}$ in simple examples.  While these estimates neglect many
possible further effects that could remove vacua, the basic picture
seems robust against effects that have been neglected to date.
Estimates of the attainable values of $W_0$ (or really $e^K |W|^2$) at
the level of the flux superpotential, fully support the assertion made
in \KKLT\ that extremely small values will be attainable.  In fact,
values much smaller than the $10^{-4}$ quoted in KKLT should arise --
the fraction of vacua with $e^K |W|^2 \leq \epsilon$ seems to fall
off only as the first power of $\epsilon$ \refs{\StatII,\unpub}.
The information available from the statistical studies is much more
detailed than we have described here (predicting for instance clustering
of vacua around the conifold \refs{\StatII,\GKT} and other interesting
dynamics), and it is heartening that the extremely complicated flux
potentials admit such simple characterizations of the statistical
properties of the resulting vacua.

However, one could still worry that although each of the ingredients in
the KKLT construction and its relatives is quite reasonable,
it might not be possible to
assemble all of the ingredients simultaneously to make a
working model.  While this
potential problem seems very unlikely from the viewpoint of low-energy
effective field theory, the issue has now been settled directly from the
string theory perspective.  In a beautiful recent paper,
Denef, Douglas and Florea (henceforth DDF) explicitly provided many
examples of Calabi-Yau orientifolds where just Euclidean D3 instantons
(effects of type 1) above)
are present in sufficient numbers to stabilize all K\"ahler moduli \DDF.
Their constructions build on earlier important work of Grassi \Grassi.
These models admit sufficiently many flux vacua that the full KKLT
construction can be carried through, as long as the existing statistical
studies are not grossly misleading.
The fraction of models where just effects of type 1) suffice to carry
out the KKLT program is not particularly small -- in the simplest
class DDF studied (fourfolds with Fano threefold base), 29 of 92
Calabi-Yau spaces could be stabilized this way.
Each such manifold leads to a rich landscape of vacua.
A subsequent work \Sethi\ also pointed to the existence of
manifolds which cannot be stabilized by such effects alone,
although this work did not address the large class, described
in \DDF, which can be stabilized.

Therefore, current evidence strongly supports the existence of a IIB
landscape as envisioned in \KKLT.  Among many issues which remain to be
clarified is the question of the extent to which effects of type 2)
above, low-energy field theory effects on D7 branes, can aid in the
stabilization of K\"ahler moduli.  In this work, we provide
some explicit compact examples where flux potentials lift D7 translation
modes, leaving a pure nonabelian ${\cal N}=1$ gauge theory on stacks of
D7s.  These will manifest gaugino condensation, although they
do not fall into the class of models studied in geometric engineering in
\Katz -- the latter work did not (and did not need to) account for D7
interactions with three-form fluxes.

Before proceeding with the technical analysis, we give a simple physical
argument which explains why one should expect the phenomena we find to
be rather generic.  The gauge theories which arise on D7 brane
worldvolumes are, in most simple examples, non-chiral gauge theories.
Even if a naive analysis at the compactification scale indicates the
presence of matter in the theory (the relevant KK scale analysis is
described in \Katz\ and other references), any further interactions at a
lower scale can therefore give the matter a mass, leaving behind a pure
gauge theory.  In these systems, there is a clear source of such further
perturbations --
the presence of $G_3$ flux.
Physically, it is then not surprising that one will often find pure
gauge theories after including the effects of flux.  We provide
explicit examples on the fourfold $K3 \times K3$ (the orientifold
$K3 \times T^2/Z_2$ in IIB language) in the following;
this example was chosen
because it is one of the simplest models that includes D7 branes,
and was analyzed in great detail in \TT\ (see also \Moore).
For the physical reason we explained above and the mathematical reasons
we explain below, we also suspect this phenomenon occurs frequently in
more complicated models, and will greatly enrich the class of
models described so far \DDF\ where the KKLT construction can be
realized.

The organization of this note is as follows.  In \S2, we review
the conditions for supersymmetry in M-theory and F-theory
compactification on a Calabi-Yau fourfold.  In \S3, we describe what
these conditions imply for $K3 \times K3$ models in more detail.  In
\S4, we provide some examples of solutions to these conditions which
yield 4d models with the D7 branes locked on loci of enhanced
non-abelian gauge symmetry.  In \S5, we explore an  orientifold example in
detail and calculate the resulting non-perturbative superpotential.
In \S6, we discuss the kinds of divisors which arise in these
singularities. In particular, there are no divisors of arithmetic genus
1 in the $K3 \times K3$ examples, although
the relevant divisors $D_i$ do satisfy $\chi(D_i,{\cal
O}(D_i)) > 1$ (and in more general examples, would have $\chi \geq 1$).
We also explain how these observations, when correctly generalized to
other Calabi-Yau fourfolds, could relax some of the conditions
stated in \refs{\Sethi,\DDF}.  In particular, we briefly discuss 
the special case where the IIB theory has a single K\"ahler
modulus, and argue that $\chi \geq 1$ divisors of the relevant
type can arise there. 
We close in \S7 with a discussion of future directions.
Some relevant details about the geometry of elliptic $K3$s and the $E_8$
lattice are relegated to appendices, as is a discussion of how
one can microscopically understand the relaxation of the
arithmetic genus 1 condition \Witten\ in the presence of flux.

The $D7/D3$ moduli in the $K3 \times T^2/Z_2$ orientifold limit  were studied  earlier in \Ferrara\
using the techniques of gauged supergravity. In particular it was shown  that
$D7$-brane moduli acquire a mass,  in agreement with our discussion here\foot{Masses for $D7/D3$ moduli have been extracted from string scattering amplitudes for type IIb orientifolds on $T^6/({\bf Z}_2\times {\bf Z}_2)$ in \LustFI.}.
Many of the solutions found below were also known to Greg Moore, from
considerations similar to those he described in \Moore.
A complementary approach to deriving the flux-induced potentials
for D7-brane moduli is described in the papers \refs{\CascalesQP,\CamaraJJ},
and we thank those authors for informing us of their results
prior to publication .

\newsec{General Conditions For Supersymmetry}

We are interested in F-theory compactifications on $K3 \times K3$ with
flux. In particular we are interested in asking whether the flux can
stabilise all the complex structure moduli (including the 7-brane
moduli) at points of enhanced gauge symmetry. In the discussion below we
take the second $K3$ to be elliptically fibered
and refer to it as $K3_2$, the other is denoted as $K3_1$ and need not
admit an elliptic fibration.

We will explore this issue by starting first in $M$ theory on $K3_1
\times K3_2$ in the presence of $G_4$ flux. This is dual to F-theory on
$K3 \times K3 \times S^1$.
We will be mainly interested in Lorentz invariant $3+1$ dimensional
solutions in  F-theory.  These are obtained in the standard fashion by
taking the size of the fiber torus in $K3_2$ to zero.

For ${\cal N}=1$ supersymmetry (four supercharges) $G_4$ must be of type
$(2,2)$. In addition $G_4$ must be primitive.
For Lorentz invariant $3+1$ dimensional solutions
$G_4$ must have one leg along the base $P^1$ and another leg along the
fiber of $K3_2$. Primitivity then reduces to the condition
\eqn\prima{G_4 \wedge J_1=0,}
where $J_1$ is the K\"ahler form of $K3_1$.

An elliptically fibered $K3$ can be described by an equation of
Weierstra\ss\ form: \eqn\wform{y^2 = x^3 + f_8(z) x + g_{12}(z),}
where $f_8(z)$ and $g_{12}(z)$ are polynomials of degree 8 and 12
respectively. This gives rise to 18 (complex) moduli which describe the
complex structure of an elliptically fibered $K3$ surface.
It is well known that the  singularities which can occur in this
equation are of A-D-E type. At a singularity the symmetry is enhanced to
the corresponding A-D-E gauge group.

We will show in specific examples that for appropriate fluxes all the
conditions of ${\cal N}=1$ supersymmetry are met and all the complex
structure moduli are stabilised such that the elliptically fibered $K3$
is at a singularity resulting in an enhanced gauge symmetry.

For this purpose it is worth discussing the above conditions in some
more detail. Some useful reference for the discussion below are \Aspin,
\GKP, \DRS, and \TT.

\newsec{The Conditions in More Detail}

$H^2(K3,R)$ is a $22$ dimensional vector space.
An inner product can be defined on this space, given by $(v_1,v_2)\equiv
\int_{K3} v_1 \wedge v_2$. This has signature $(3,19)$.
$H^2(K3,Z)$ can be thought of as a lattice, $\Gamma^{3,19}$, embedded in
this vector space. In a suitable basis the inner product for the basis
elements of this lattice takes the form of the matrix $$U\oplus U \oplus
U \oplus (-E_8) \oplus (-E_8),$$
where $U$ is the $2 \times 2$ matrix
\eqn\defu{
U = \left(\matrix{0 & 1 \cr 1 &0}\right)}
and $E_8$ is the Cartan matrix for the $E_8$ lattice.

The holomorphic two-form $\Omega$ on $K3$ is given by  a spacelike
oriented two-plane in $H^2(K3,R)$. The moduli space of complex
structures
then corresponds to the space of all such distinct two-planes.
Up to discrete identifications this is
$$G=O(3,19)/O(2) \times O(1,19),$$
which is $40$ dimensional. In the discussion below we will refer to both
the holomorphic two-form and the associated space-like two plane as
$\Omega$.
Clearly, all two-forms in $H^{(1,1)}$  must be orthogonal to $\Omega$.

For the $K3$ surface to be elliptically fibered, a sublattice $U\subset
\Gamma^{(3,19)}$ must be orthogonal to  the two-plane \foot{More
accurately the sublattice has two basis elements whose inner product
takes the form $U$, \defu. In an abuse of notation we will refer to the
sublattice itself as $U$ below. } $\Omega$. The moduli space of complex
structures for elliptically fibered $K3$s is then given (again up to
discrete identifications) by
$O(2,18)/O(2) \times O(18)$. This is  $18$ (complex) dimensional, the
counting agrees with the moduli in the Weierstra\ss\ form, \wform.
The Picard lattice of $K3$ is defined as the lattice of integral
two-forms which are of $(1,1)$ type. That is
$$Pic(K3)=H^2(K3,Z) \cap H^{(1,1)}(K3). $$
So we see  that for the $K3$ to be elliptically fibered
$U\subset Pic(K3)$.

How this  requirement for an elliptic fibration  comes about will be
discussed further in the Appendix. For now we simply note that the
two-forms dual to  the  base $P^1$ and the fibre torus both lie in $U$.
We saw above that $G_4$ must have one leg along the base and one along
the fibre of the elliptically fibered $K3$. This means that $G_4$ cannot
contain any element in $U$ and so must be orthogonal to $U$.
That is for any $u \in U, G_4\cdot u=0$.

We can now restate the conditions for supersymmetry as follows.
$G_4$ must be chosen to have two legs along $K3_1$ and two legs along
$K3_2$. A complex structure on $K3_1$ and $K3_2$ must exist such that
$G_4$ is of type $(2,2)$. The resulting  Picard lattice for elliptically
fibered $K3_2$ must contain the sublattice $U$. $G_4$ must  be
orthogonal to $U$. And finally, the K\"ahler form of $K3_1$ must satisfy
the primitivity condition, \prima, which can be restated in terms of the
inner product defined above as $J_1\cdot G_4=0$.

Let us now discuss the singularities in the Weierstra\ss\ form, \wform,
in some more detail. A root of $\Gamma^{3,19}$ is defined to be a vector
$\alpha \in  \Gamma^{3,19}$ with $\alpha \cdot \alpha=-2$. A singularity
in the Weierstra\ss\ form arises if there is a
 root of $\Gamma^{3,19}$ which lies in $Pic(K3)$ and which is
orthogonal to $U$. That is, if there is a root orthogonal to both $U$
and $\Omega$.\foot{The requirement that the  Einstein
metric on  K3 is at an orbifold singularity is somewhat different. The
metric corresponds to a choice of space-like three-plane and an orbifold
singularity occurs if a root is orthogonal to this three-plane. This
ensures that orbifold has not been resolved either by K\"ahler
deformations  or complex structure deformations. For F-theory we are
only interested in the complex structure deformations which preserve the
elliptic form
 and we therefore require  orthogonality with respect to
$\Omega$ and $U$.} The orthogonal roots form the root  lattice of
an A-D-E algebra. The singularity is of the corresponding A-D-E type.

Before proceeding, let us make the following two comments.
First, it is worth briefly recapitulating  why one expects all complex
structure moduli to be generically stabilised in an ${\cal N}=1$ susy
solution. The requirement that $G_4$ is of type $(2,2)$ means that the
$(4,0),(0,4)$ and $(1,3), (3,1)$ components in $G_4$ must vanish. This
imposes one more condition than the number of complex structure moduli.
 For a choice of flux where a susy solution does exist this implies
that all the complex structure
moduli should generically be lifted.

Second,
the four-form flux, $G_4$, gives rise to three form flux in the IIB
description. Let the holomorphic and anti-holomorphic differentials
along the elliptically fibered torus of $K3_2$ be $dz$ and $d{\bar z}$,
and
$\phi$ be the modular parameter (the axion-dilaton in the IIB theory).
Then $G_4$ can be expressed in terms of the three-form flux in IIB,
$G_3=F_3-\phi H_3$, as follows:
$$G_4=-{1 \over \phi-{\bar \phi}}G_3\wedge d{\bar z} + {1\over \phi
-{\bar \phi}}{\bar G_3}\wedge dz.$$ $G_4$ can also give rise to two-form
flux, $F_2$,  in the  world-volume theory of  the $D7$ branes. For
example, if two $D7$-branes come together giving rise to an $A_1$
singularity and if $\alpha$ is the corresponding root of $\Gamma^{3,19}$
that is orthogonal to $\Omega$, then a non-trivial $F_2$ is turned on in
the relative $U(1)$ between the two $D7$-branes if $G_4$ has a component
of the form, $G_4=\beta \wedge \alpha$, where $\beta$ is an integral
two-form in $K3_1$. Note however that if $Pic(K3_1)$ is trivial - as
will be the case generically -
 such a component is not
allowed by supersymmetry.  This follows from noting that since $\alpha$
is orthogonal to $\Omega$ it must be of type $(1,1)$. Supersymmetry
requires that  $G_4$ is of type $(2,2)$, this means $\beta$ must be  an
$(1,1)$ form in $K3_1$ and must therefore belong to $Pic(K3_1)$.
In any event, we shall avoid turning on such $F_2$ fluxes in our
constructions.

\newsec{Examples}

\subsec{Simplest examples}

We will now construct an explicit example where all the complex
structure moduli are stabilised at a point of enhanced symmetry.

Consider the six-dimensional subspace $H_{3,3}= U \oplus U \oplus U$ of
$H^2(K3,Z)$. In a suitable basis, which we call $(e_1, \cdots e_6)$  the
 inner product in this subspace
 takes the form,
$2 \eta_{3,3}$,
with $\eta_{3,3}=diag(1,1,1,-1,-1,-1)$. The basis elements are chosen so
that $e_1,e_4$ span the first $U$ sublattice and so on. Also we note
that
 $e_1,e_2,e_3$ are space-like and the rest are time-like.

Now consider the flux
\eqn\ega{{G_4 \over 2 \pi}=e_1\wedge {\tilde e_1} + e_2 \wedge {\tilde
e_2}} where $e_1,e_2$ and ${\tilde e}_1, {\tilde e}_2$  refer to vectors
in the integral lattice of $K3_1$ and  $K3_2$ respectively.
This flux satisfies the requirement of having two legs along the two K3s
respectively.

It is easy to see that $G_4$ can be written as
\eqn\egb{{G_4 \over 2 \pi} = {1 \over 2}\left[(e_1 + i e_2) \wedge
({\tilde e}_1 - i {\tilde e}_2)
       +(e_1 - i e_2) \wedge ({\tilde e}_1 + i {\tilde e}_2) \right]  }
So by choosing the complex structure $\Omega_{1}, \Omega_2$ of
$K3_1,K3_2$ as follows, \eqn\csto{ \Omega_1 =  (e_1 + i e_2)}
\eqn\cstt{ \Omega_2 =  ({\tilde e}_1 + i {\tilde e}_2)~,}
we see that
\eqn\egc{{G_4 \over 2 \pi} ={1 \over 2}\left[\Omega_1 \wedge
{\bar\Omega}_2 + {\bar\Omega}_1 \wedge \Omega_2\right],}
and is therefore of type $(2,2)$.
We note that the identification \csto, \cstt\ is consistent with
requiring that $\Omega_{1,2}$ are space-like two-planes in
$H^2(K3_{1,2},R)$. Also, since, $e_{1}\cdot e_1=e_2\cdot e_2, e_1\cdot
e_2=0$ and similarly for ${\tilde e}_{1,2}$, \csto, \cstt\  are consistent
with the requirements that
  $\Omega_{1} \cdot \Omega_1=\Omega_2\cdot \Omega_2=0$.
In addition note that ${\tilde e_3}, {\tilde e_6}$ span a subspace $U$ of
$H^2(K3,Z)$ and are orthogonal to $\Omega_2$. This ensures that $K3_2$
is elliptically fibered.

 $G_4$ is orthogonal to ${\tilde e_3}, {\tilde e_6}$ and therefore to
$U$,
this ensures that  one leg of $G_4$ is along the base and the other
along the fiber of $K3_2$.

Finally a K\"ahler form for $K3_1$ can be chosen meeting the condition
\prima. The K\"ahler form corresponds to a space-like direction in
$H^2(K3,R)$ orthogonal to $\Omega_1$. In the example above we could take
this direction to be along $e_3$, then we see that $e_3 \cdot G_4=0$ so
that the condition of primitivity is met.

Thus we see that all the requirements for an ${\cal N}=1$ solution are
met in this example.

We will argue next that the complex structure moduli are all frozen
about this point. Consider a small deformation in the complex structure
of $K3_1$. Under it $\Omega_1 \rightarrow  \Omega_1 + \chi_1$ where
$\chi_1$ is a $(1,1)$ form in $K3_1$. Similarly $\Omega_2 \rightarrow
\Omega_2 +  \chi_2$. Under this transformation,
$$ {G_4 \over 2 \pi} = {1 \over 2}\left[\Omega_1 \wedge {\bar\Omega}_2 +
{\bar\Omega}_1 \wedge \Omega_2 + \chi_1 \wedge {\bar\chi}_2+
{\bar\chi}_1 \wedge \chi_2 + K_4\right]$$ with the four form $K_4$
defined as
$$ K_4 = \chi_1 \wedge {\bar\Omega}_2 + {\bar\chi}_1 \wedge \Omega_2
       + \chi_2 \wedge {\bar\Omega}_1 + {\bar\chi}_2 \wedge \Omega_1. $$
Now since, $\Omega_{1,2}, {\bar \Omega}_{1,2}$ are linearly independent,
$K_4$ cannot vanish, so $G_4$ cannot be of type $(2,2)$ after the small
deformation.
 Thus for small deformations, $G_4$ no longer remains
$(2,2)$ and the complex structure moduli are all lifted.

Let us now turn to describing the enhanced symmetry.
We note that the lattice vectors of $E_8 \oplus  E_8 \subset
\Gamma^{3,19}$ are  orthogonal to $\Omega_2$.  In addition they are
orthogonal to the $U$ sublattice spanned by $({\tilde e_3}, {\tilde
e_6})$. Similarly the vectors ${\tilde e}_4,{\tilde e}_5$ are roots
which are orthogonal to both $ \Omega_2$  and $U$.
Thus the enhanced gauge symmetry in this example is $SU(2) \times SU(2)
\times E_8 \times E_8$.

Finally we note that the membrane tadpole condition is met in M-theory.
Since $${1 \over 2} \int {G \over 2 \pi} \wedge {G \over 2 \pi} = 4 <
{\chi \over 24} =24,$$ one will need to add $20$ M2 branes (D3 branes)
in M-theory (F-theory).

\subsec{An Orientifold example}
By starting with the example above and changing the flux one can alter
the complex structure $ \Omega_2$ so that the gauge symmetry is reduced.
In particular the symmetry can be broken to
$SO(8)^4$. The resulting model then corresponds to taking the
elliptically fibered $K3_2$ at the orientifold point,
with $4$ D7-branes at each orientifold plane. It is worth examining this
orientifold limit of the example above  in more detail.
This will allow us to explicitly calculate the masses of the D7-branes.
It will also allow us to make contact with the discussion in \TT. We
will find below that the orientifold examples corresponds to solutions
of the type (2+,0-) in the classification of \TT\ (section 3.3).

It is quite straightforward to find flux that will stabilise the complex
structure at an orientifold  singularity. The complex structure moduli
space of the elliptically fibered $K3_2$ is 18
 dimensional. At an orientifold point $16$ of these moduli correspond to
the location of D7-branes along
the $T^2$ base of $K3_2$ and the remaining two moduli are the
dilaton-axion and the complex structure of the $T^2$. Requiring that the
complex structure is at a $(D_4)^4$ singularity fixes the locations of
the D7-branes  while
allowing the other two moduli to vary.
We will proceed in two steps in the discussion below, first finding a
particular point in the complex structure moduli space
where the singularity is of $(D_4)^4$ type, and then determining a flux
which fixes the complex structure at this point.

The model under consideration
is dual to the heterotic string  on $K3 \times T^2$. It is well known
that by turning
 on Wilson lines on the heterotic side
one can break the gauge symmetry down to $SO(8)^4$.
Using the duality map,   one can   then map this to a location
 in the complex structure  moduli space of  $K3_2$.
Before proceeding let us note that, upto a sign convention, our discussion of
 Wilson lines in the heterotic string  will be based on \Polchinski.

For simplicity we take the heterotic theory with a square $T^2$ at the
self-dual point, with no $B$ field  and
with appropriate Wilson lines turned on. The resulting complex structure
of $K3_2$  can then  be described as follows.
The complex structure of $K3_2$ corresponds to a two-plane $\Omega_2$.
In \S4.1\  we described the basis vectors $e_1, \cdots e_6$ of   the
subspace $H_{3,3} \subset \Gamma^{3,19}$, with $e_1,e_4$ spanning the
first $U$ subspace etc. It is easy to see that $n_1={e_1+e_4 \over 2}$
is a null vector, $n_1 \cdot n_1=0$, meeting the
 condition $n_1\cdot e_1=1$. Furthermore it is a basis element for the
$U$ sublattice of $\Gamma^{3,19}$. Similarly, we define the null
vector, $n_2={e_2+e_5\over 2}$,
which is a lattice basis element for the second $U$ sublattice of
$H_{3,3}$. We will also need to introduce  a basis in the $(-E_8) \oplus
(-E_8) $ sublattice of $\Gamma^{3,19}$.
This is done by choosing vectors, $E_I, I = 1 \cdots 16, E_I \cdot
E_J=-\delta_{IJ}$. The roots of $(-E_8) \oplus (-E_8)$  are then given
by $q^IE_I$ for suitably chosen $q^I$ as discussed in
Appendix B.
The required  spacelike two plane corresponds to a choice of two
linearly independent space-like  vectors. These are given by ${\hat
{\tilde e}}_1, {\hat{\tilde e}}_2$ respectively, where
\eqn\defeto{\eto= {\tilde e}_1 + W_I E_I + {{\tilde n}_1 \over 2} W_I
W_I} \eqn\defett{\ett ={\tilde e}_2 + {\tilde W}_I E_I + {{\tilde n_2}
\over 2} {\tilde W_I}
 {\tilde W_I}}
Here $W_I, {\tilde W_I}$ denote the two Wilson lines. Also as in the
previous section, we are following conventions where
 the tilde superscript as in  ${\tilde e_1}$ etc, refers
to elements of $H^2(K3_2), \Gamma^{3,19}(K3_2)$.
With
\eqn\wa{W_I=diag(1,0^7,1,0^7)}
\eqn\wb{{\tilde W_I}=diag(0^4, {1\over 2}^4, 0^4,  {1\over 2}^4),} one
can show that the  roots of $\Gamma^{3,19}$ orthogonal to $\eto$, $\ett$
correspond to the gauge group $SO(8)^4$.
It is easy to see that $\eto, \ett$ are linearly independent and
therefore define a space-like two-plane. Identifying this two-plane with
$\Omega_2$ gives the location of a $(D_4)^4$ singularity in
the complex structure moduli space of $K3_2$.

We now turn to determining the required flux which will stabilise the
complex structure at this point in moduli space.
As discussed in appendix B,  $2 \eto, 2 \ett$, are  elements
of the integral lattice, $\Gamma^{3,19}$.  So we can  consider turning
on the four-form flux, \eqn\altfour{G_4= 2 e_1 \wedge \eto + 2 e_2
\wedge \ett.}
Since $\eto, \ett$ satisfy the relations $\eto\cdot \eto=2, \ett\cdot
\ett=2, \eto\cdot \ett=0$ , we see that the discussion
in the previous  section  goes through  unchanged  showing that a
complex structure for $K3_1 \times K3_2$ exists given by \egc, with
$G_4$ being  of type $(2,2)$. $\Omega_1$ is unchanged from \csto, and
$\Omega_2$ is given by
\eqn\newcst{\Omega_2=\eto+i \ett.}

Let us end this section with a few comments.
First, since $G_4$ is of form \egc,  our discussion in the previous
section still goes through leading to the conclusion that the complex
structure moduli are all stabilised. Second, we note that $G_4$ above
does not have any component along the root lattice of $SO(8)^4$, thus no
gauge field flux $F_2$ is turned on in this case along the 7-brane world
volumes. Third,  the total contribution to the membrane tadpole
condition is $16$, this means $8$ $D3$ branes would have to be added in
the F-theory description. Fourth,  the fact that this example
maps to the $(2+,0-)$ case in \TT\ as mentioned above,
 follows simply by noting that  there are only two linearly independent
two-forms,
$e_1,e_2$ of $K3_1$ in $G_4$.
Finally, for simplicity here we have focused on one choice of flux. More
generally other choices of flux can also stabilise the complex structure
at a $(D_4)^4$ singularity
 for other values of the dilaton-axion and $\tau$.


\newsec{The orientifold model in more detail}
It is worth exploring the orientifold model above  in some more detail.
We have given a general argument above that all the complex structure
moduli are stabilised. Here we would like to explicitly verify this for
the D7 moduli by calculating their mass. This will allow us to calculate
in the next section
the leading contribution (at large volume) to the
non-perturbative superpotential  due to gaugino condensation.

\subsec{D7-moduli mass}
The complex structure moduli space of the elliptically fibered $K3_2$ is
$18$ complex dimensional. At the orientifold point, $16$ of these $18$
moduli correspond to Wilson lines that give the location of the
$D7$-branes
along the base $T^2$ of $K3_2$. The remaining two moduli
correspond to the dilaton-axion, and the modular parameter
of the base $T^2$. At the orientifold point, as was mentioned above, $4$
D7 branes are located at each O7-plane.  By symmetry one can see that
all the $D7$-branes must have the same mass. Displacing one $D7$-brane
from the O7-plane breaks the symmetry to $SO(6)\times SO(8)^3$. The
corresponding Wilson lines are given by
\eqn\wan{W_I=diag (1, \alpha, 0^6, 1, 0^7),}
\eqn\wbn{{\tilde W_I}=diag(0,\beta, 0^2, {1 \over 2}^4, 0^4,  {1\over
2}^4),}
 where $\alpha, \beta$ are the locations of the
$D7$-brane along the $T^2$. The resulting complex structure of $K3_2$ is
given by \newcst, where $\eto,\ett$ are now given by
\eqn\neweto{\eto={\tilde e_1}+W_IE_I+{{\tilde n_1}\over 2} W_IW_I +
{{\tilde n_2} \over 2} W_I {\tilde W_I},}
\eqn\newett{\ett={\tilde e_2}+{\tilde W_I} E_I  +{{\tilde n_2}\over
2}{\tilde W_I} {\tilde W_I} +{{\tilde n_1} \over 2} W_I {\tilde W_I}, }
 with the Wilson lines \wan, \wbn.
As a check it is easy to see that this complex structure corresponds to
an unbroken $SO(6)\times SO(8)^3$ symmetry.

We now turn to determining the mass for the D7-brane moduli.
The   superpotential is given by
\eqn\superpot{W=\int G_4 \wedge \Omega_4.}
$G_4$ is fixed and given by \altfour,  with $\eto, \ett$ being given by
\defeto, \defett,
 with the Wilson lines, \wa, \wb, respectively.
$\Omega_4=\Omega_1\wedge \Omega_2$.
As the 7-brane moves away from the O7 plane
the complex structure of $K3_2$, $\Omega_2$, changes, as described in
the previous paragraph. $\Omega_1$ is fixed and given by \csto.
By substituting these expressions in $W$ above it is straightforward to
see that it takes the form, \eqn\fsp{W=4 (\alpha+ i \beta)^2.}
We saw above that the complex scalar $\alpha+i\beta$ is the location of
the D7-brane along the $T^2$ base of $K3_2$. We will denote it by
$\Phi=\alpha+i \beta$ in the discussion below. The quadratic term in the
superpotential shows that this modulus acquires a mass. From \superpot,
it is clear that the mass is linear in the flux.

One final comment.
Our result above for the 7-brane moduli mass agrees with an earlier calculation
using the methods of gauged supergravity, \Ferrara.

\subsec{The Non-perturbative Superpotential}
The low-energy dynamics of the orientifold model discussed above is that
of an ${\cal N}=1$ $SO(8)^4$ gauge
theory. The gauge group arises form the gauge fields on the D7-branes.
The D7-branes wrap $K3_1$ and are transverse to the $T^2$ base of
$K3_2$.
 The D7-brane locations on the $T^2$ are  moduli which are  adjoint
chiral superfields in the gauge theory. There is one adjoint field for
each $SO(8)$ group.
We saw above that these moduli   acquire a mass
due to the flux.  This mass scales with the radius of compactification,
$R$, as $1/R^3$. At energy scales below this mass scale the low-energy
dynamics is that of a pure ${\cal N}=1$ SYM theory with gauge group
$SO(8)^4$. It is well known that a non-perturbative superpotential is
generated in pure SYM theory due to gaugino condensation. The form of
this superpotential can be determined by standard field theory
techniques.
 Since the gauge fields arise from the D7-brane world volume, the gauge
coupling
is given by
$$S={8 \pi^2 \over g_{YM}^2} + i \theta = e^{4u -\phi} +i b, $$
where $e^{4u}$, is the volume of $K3_1$,   $\phi $ is the dilaton and
$b$ is an axion that arises from the RR four-form. The  gauge couplings
of the four $SO(8)$ groups are the same.

Standard field theory techniques then show that the non-perturbative
superpotential is given by :
\eqn\wnp{W_{NP}=A e^{-S/c_2}m}
where $m$ is the mass of the adjoint chiral superfield.
$A$ above is a coefficient that depends  on the expectation values of
the frozen
 complex structure moduli, and
 $c_2$ is the dual Coxeter number
in the adjoint representation of the gauge group.
For $SO(2n)$,  $c_2=(2n-2)$, giving in particular  $c_2=6$ for $SO(8)$.

Let us briefly sketch how this result is obtained.
We  denote the tree-level superpotential, \fsp, as $W_{tree}$ in the
discussion below. This  tree-level
superpotential has a $U(1)$ R-symmetry under which $\Phi \rightarrow
\Phi e^{i\theta}.$ The R-symmetry corresponds to rotations in the plane
of the $T^2$ in $K3 \times T^2/Z_2$. We see that $W_{tree}$ has charge
$2$ under this R-symmetry.
This symmetry is anomalous in the quantum theory and it is easy to see
that $S \rightarrow S - 2 i c_2 \theta$ under it.
Now  $W_{NP}$ must transform in the same way as $W_{tree}$ under this
symmetry, this fixes the dependence on $S$ in \wnp.
Similarly $W_{tree}$ has an R-symmetry
under which $m$ transforms with charge $2$ and $\Phi$ is invariant. The
R-symmetry can be shown to be non-anomalous so $S$ is invariant under it.
 This determines the $m$ dependence in \wnp \foot{
These two R symmetries act on the GVW superpotential as follows. Under
the first symmetry, $\Omega \rightarrow \Omega e^{2i\theta},$ and the
flux
$G_4$ is invariant. Under the second symmetry, $G_4 \rightarrow G_4
e^{2i\theta}$, and $\Omega$ is
 invariant.
Towards the end of section 6 we will refer to a $U(1)$ discussed in
\Witten . This symmetry is a combination of the two R-symmetries
discussed above. Under it, with our choice of normalisation, $\Phi$ and
$m$ have charges $2,-2,$ respectively,
while $\Omega, G_4,$ have charges $4,-2$.}.

The non-perturbative superpotential \wnp\  results
in a potential for the  volume modulus
of $K3_1$. In general one would expect that  \wnp\ is  not exact in
string theory and there are corrections to it which are subleading at
large volume. Requiring that $W_{NP}$ transforms correctly under full
duality group, together with the large volume dependence determined
above,  might help fix the form of these subleading corrections.

\newsec{Divisors contributing to the superpotential}

Let us begin by reminding the reader of Witten's argument \Witten\
that divisors $D$ of arithmetic genus, $\chi= 1$,  are the relevant ones for
superpotential generation in M-theory
compactification on a Calabi-Yau fourfold $X$.  For a Euclidean brane
instanton to contribute to the superpotential, it should break half
of the space-time supersymmetry, leaving precisely two fermion zero
modes. For a Euclidean M5 brane instanton the zero modes are determined by the
$h^{k,0}(D)$ cohomology groups of the divisor wrapped by the Euclidean 5-brane.
Witten considered the $U(1)$ symmetry of the normal bundle
$N$ to $D$ and argued that modes arising from the cohomology groups $h^{2p,0}(D)$ and
$h^{2p+1,0}(D)$ have opposite $U(1)$ charges and can pair up. The index
obtained after grading the zero modes by the sign of their charge under this $U(1)$ symmetry is
$2 \chi$. Thus a necessary (although not always sufficient) 
condition for the non-perturbative
superpotential to arise is that $\chi=1$.
We will see that after
accounting for the effects of G-flux, this conclusion is
modified, and in general $\chi > 1$ divisors can also contribute
to the space-time superpotential; the loophole was anticipated on
p.10 of \Witten.

\subsec{In the $K3 \times K3$ examples}

Here, we describe in slightly more detail the geometry of the singular
$K3 \times K3$ compactifications that must be yielding our gaugino
condensates.  In each of these cases, the singular elliptic $K3_2$ has
an A-D-E singularity. For simplicity we will focus on the case of
$A_{N-1}$ here, but very similar remarks apply to the other two cases.
Our discussion of the relevant geometries follows \Katz.

We have an $A_{N-1}$ singularity over $K3_1$.  Denote the exceptional
divisor by $D$.  $D$ is the union of $N-1$ irreducible components
$D = \cup_i D_i$.
After blowing up, the fiber over $K3_1$ will consist of
$N$ $P^1$s, intersecting in such a way as to form the
affine Dynkin diagram for $SU(N)$ (the $D_i$ have been supplemented by
an additional divisor $D^\prime$, which is the closure of the
complement of the exceptional set $D$ inside the resolved elliptic
fiber) \foot{The additional $P^1$ which gives rise to the divisor
$D^{\prime}$ can be understood as follows. Since $K3_2$ is elliptically
fibered there is a null vector, $n$, dual to the $T^2$ fiber. Like the
roots of $A_{N-1}$, $e_i$,
$n$ is also  orthogonal to the the two-plane, $\Omega_2$.
The additional $P^1$ is dual to
 $n-\sum e_i$. And $e_i$ together with $n-\sum_ie_i$
  give rise to the Dynkin diagram of affine $A_{N-1}$.}.

In general examples that would arise in elliptic Calabi-Yau fourfolds,
this story would generalize as follows.  One would look for
singularities of the elliptic fibration of Kodaira $I_{N}$ type
over some surface $S$ in the base $B$ (the singularities may worsen to
$I_{N+1}$ at codimension one in $S$, etc.).
The $D_i$ (and $D^\prime$) are then nontrivial $P^1$ bundles over
the surface $S$ wrapped by the D7 branes $\pi:D_i \to S$.
Our particular example is however quite simple: $S$ is $K3_1$ and the
fiber is constant along $S$.  Therefore, the $D_i$ simply take the form
$K3_1 \times P^1$.  It is easy to see that the arithmetic genus of these
cycles satisfies
\eqn\arithgen{\chi(D_i,{\cal O}(D_i)) = h^{0,0}(K3) + h^{2,0}(K3) = 2~.}
Hence, while it is completely clear from the 4d and 10d perspectives
that our examples have pure Yang-Mills sectors which will undergo
gaugino condensation, there need not be divisors of arithmetic genus 1.
This is in keeping with the remarks in \Witten\ about how infrared gauge
theory effects may not give the correct naive zero mode count required
for superpotential generation.

It is suggestive that an M5-brane wrapping $D$ (i.e., all $N$ $P^1$s)
would be wrapping a cycle of $\chi(D) = 2N$.  This coincides with the
number of fermion zero modes which are present in a naive instanton
calculation in pure $SU(N)$ ${\cal N}=1$ Yang-Mills theory, where
the gaugino condensate $\langle \lambda \lambda \rangle$
scales like the $N$th root of a gaugino 2N-point correlator (see e.g.
\meurice, pages 184-186).
This underscores once again the fact that gaugino condensation is not an
instanton effect
in the 4d field theory picture, and it is not very surprising
that there is no instanton with two fermion zero modes.

\subsec{More general cases}

More generally, our construction suggests the following.  Consider any
case where in a fourfold, there is a Kodaira type $I_{N}$ degeneration
over $S$ as described above (in physics language, this is the situation
when there are $N$ D7 branes wrapping $S$).  In general there
will be $I_{N+1}$
curves in $S$ also (physically, these are curves where another D7
intersects the stack of $N$ D7s wrapping $S$).
One can then consider an M5 brane wrapping the cycles
$D_i \to S$ and $D^\prime \to S$, which are fibrations over $S$ with
$P^1$ fibers.
Using the Leray spectral sequence as on p.6 of \Katz, we see
the following.  Since the $P^1$ fiber has $h^{1,0}(P^1)=0$, and the only
holomorphic function on $P^1$ is a constant,
$H^{i,0}(D_i) \simeq H^{i,0}(S)$.
Therefore, these cycles will have arithmetic genus
\eqn\aritn{\chi(D_i) = h^{0,0}(S) - h^{1,0}(S) + h^{2,0}(S)~.}
Let us assume for simplicity that $S$ is simply connected, so
$h^{1,0}(S) = 0$. This is not a serious restriction, as in many cases
$S$ inherits its first cohomology from the cohomology of the base of the
F-theory elliptic fourfold (by the Lefschetz hyperplane theorem),
and for simple examples this vanishes.
We then see that $\chi(D_i) > 0$, and all zero modes but the one
arising from $h^{0,0}(S)$ are in correspondence to adjoint matter fields
on the D7 branes wrapping $S$.  (Recall a D7 brane wrapping $S$ receives
$h^{1,0}$ adjoint fields from Wilson lines on $S$, which vanishes
for us, and $h^{2,0}$ adjoint
fields from deformations of $S$ inside the compactification manifold).\foot{In
general there can also be fundamental matter fields arising from the
intersections with other D7s.  These also yield a nonchiral spectrum
in simple examples, and analogues of the phenomena
we are exploring here should
be expected to occur.  Some examples of $SU(N_c)$ theories
with quark flavors arising from D7s are described in \Diaconescu.
In many cases, more exotic theories which are not yet well understood
can also arise.}
Hence in all cases where the matter fields are lifted by three-form
fluxes leaving an $I_{N}$ degeneration over $S$, there are
also cycles present of the appropriate arithmetic genus
($\chi \geq 1$) to possibly contribute to the superpotential.

We therefore expect that in many examples of IIB compactifications with
flux, one will obtain contributions to the nonperturbative
superpotential from stabilized coincident D7 branes.  This
can happen sometimes even in the absence
of cycles of arithmetic genus 1 in the related Calabi-Yau fourfold.
However in all such cases, we expect (as in the $K3 \times K3$ examples)
that divisors of $\chi \geq 1$ exist.

We see no obstruction to such examples arising even in cases where the
IIB compactification manifold has a single K\"ahler modulus.
For instance, it is easy to write
down examples in the elliptic fibration over $P^3$
which have singularities of various Kodaira types
over a surface $S$ (which has $h^{1,0}(S)=0$)
in $P^3$.
The question is then whether
appropriate fluxes can stabilize D7s on such a locus.
If so, the
resulting theory would exhibit gaugino condensation and a
nonperturbative superpotential for the K\"ahler modulus.
Our explicit examples of this in $K3 \times K3$ give us confidence that
the phenomenon will happen in more general examples.

A recent paper \Sethi\ argued that vertical $\chi=1$ divisors
cannot appear in elliptic fourfolds $X \to B$ with $h^{1,1}(B)=1$,
and that hence such models cannot have nonperturbative superpotentials
for the K\"ahler modulus.  Since 
our comments suggest otherwise, let us address the contradiction.
The arguments presented in \Sethi\ do not prove that vertical $\chi=1$
(or $\chi>1$) divisors cannot appear in models with $h^{1,1}(B)=1$.
Equation (18) in \S2.5\ of \Sethi, for
the total Chern class of $X$, is
in general incorrect for models where $h^{1,1}(X)>2$ but $h^{1,1}(B)=1$. 
This equation plays an important role in constraining the 
possible arithmetic
genera of divisors.
The argument of \S2.6 in \Sethi, which 
shows that no base $B$ with $h^{1,1}(B)=1$
can be globally fibered over a surface $S$, is true.
However, that fact is not relevant to the phenomena under discussion here.
A global fibration of $B$ over some $S$ is not needed to obtain
nonabelian gauge symmetries from coincident D7 branes in $B$,
and it is straightforward to write down examples of 
$I_n$ singularities with $n>1$ in the fourfolds
of \Klemm\ with $h^{1,1}(B)=1$.

In a forthcoming paper \DDFK, the 
question of a non-perturbative superpotential in cases
where the IIB compactification 
has a single K\"ahler modulus will be discussed in much greater depth.
Several explicit examples of elliptic fourfolds $X \to B$ with $h^{1,1}(B)=1$ 
(and $h^{1,1}(X)>2$) and 
vertical divisors of arithmetic genus one have been found and will
be presented in \DDFK.  The simplest examples arise by
working with the elliptic fourfold in $WP^{5}_{1,1,1,1,8,12}$ 
(which is elliptically fibered over $P^3$), constructing singularities 
of various Kodaira types over a $P^2$ in the $P^3$, and blowing
them up.  

Finally, let us close this section by noting that 
KKLT discussed the single K\"ahler modulus case
solely for simplicity of exposition.  The arguments presented
there are more general and do not 
depend in any important way on this condition.

\newsec{Discussion}

There are several obvious directions for further work.
Needless to say, it would be interesting to
find explicit examples of the phenomena we have seen in $K3 \times K3$
in other Calabi-Yau fourfolds.
While explicit examples where all K\"ahler moduli can be stabilized
already exist \DDF, we believe the kinds of effects described here
will greatly broaden the class of examples.

It would also be interesting to understand the story when our models are
reduced to 3d directly in the language of M5 instantons.
A 4d ${\cal N}=1$ pure gauge theory, reduced to 3d on a circle, exhibits
 superpotential
generation due to abelian instantons on the Coulomb branch, and a
microscopic version of this
involving M5 instantons appeared in \Katz.  Our situation involves a  4d
theory which has chiral multiplet matter fields, which are then given a
mass.  After the reduction to 3d, an instanton computation
in the 3d field theory
must still yield a nonzero result: the mass terms for the matter fields
can be pulled down to absorb any extra zero modes which naively appear
in the instanton calculation.
This macroscopic phenomenon must have a microscopic analogue in M5
instanton calculations, and it would be interesting to understand this
in detail. One  comment in this regard is worth making.
As discussed in \Witten\ the arithmetic genus of the divisor wrapped by
the $M5$ brane
 corresponds to an index graded by a $U(1)$ symmetry.
This $U(1)$ symmetry is the structure group of the normal bundle of the
divisor.  In the $K3_1\times K3_2$ case, as is discussed further in appendix C,
 one finds that the  symmetry is
broken by the flux.
 This suggests that
in the M5 brane worldvolume theory, modes coming
 from $h^{2,0}$ can pair up amongst
themselves and became massive in the presence of flux, allowing for a
superpotential even when $\chi>1$.

Finally, we note that in our $K3 \times K3$ examples, $G_4$ fluxes
which reduce purely to
three-form fluxes in the IIB language (no field strengths $F_2$ turned on
in the D7 branes) suffice to stabilize all D7 brane moduli.  It
would be interesting to see if this phenomenon persists in more
generic examples.\foot{It
is clear that generic $G_4$ fluxes which reduce to both
three-form fluxes and $F_2$ fluxes in IIB, do suffice to stabilize
all D7 moduli in generic examples.  This follows from simple
counting of equations and fields given the GVW superpotential
for a CY fourfold.}

\medskip
\centerline{\bf{Acknowledgements}}
\medskip
We would like to thank P. Aspinwall, F. Denef, O. DeWolfe,
M. Douglas, N. Fakhruddin,
B. Florea, A. Grassi, N. Iizuka,  S. Katz,  G.
Moore, S. Shenker and A. Uranga for helpful discussions.
The work of S.K. was supported in part by a David and Lucile Packard
Foundation Fellowship for Science and Engineering, the National
Science Foundation, and the Department of Energy.
S.P.T. acknowledges the support of the Swarnajayanti Fellowship,
DST. Govt. of India.
The work of L.G., P.K.T. and S.P.T. was supported by the Govt. of India,
most of all they thank the people of India for enthusiastically
supporting research in string theory.

\appendix{A}{More on Elliptically Fibered $K3$}

Here we will describe how the requirement that $U\subset Pic(K3)$ arises
for an elliptically fibered  $K3$.

In general, as we know, the second cohomology group for $K3$ is given by
the lattice
$$\Gamma_{3,19} = U\oplus U \oplus U \oplus (-E_8) \oplus (-E_8)$$ where
$U$ is the hyperbolic plane, \defu,
and $E_8$ is the Cartan matrix for the $E_8$ lattice. Being an elliptic
fibration, the $K3$ obviously has, at least, two algebraic curves
embedded in it, one being a nontrivial section of the bundle (which is
guaranteed from the Weierstra\ss\ form) and the other being the elliptic
fiber. Obviously, they are elements of the second homology group
$H_2(K3,Z)$ (which is same as $H^2(K3,Z)$), and since they
are algebraic
they also belong to $H^{1,1}(K3)$. Thus they belong to the Picard
lattice of $K3$.

We can obtain the intersection matrix for these two curves as follows.
In general a genus $g$ curve in $K3$ has self-intersection number equal
to $(2 g - 2)$ and hence for the section (which is a $P^1$) it is $- 2$,
whereas for the fiber it is zero. Since both the curves intersect each
other transversely, their intersection number is one. Thus the
intersection matrix is
$$\left(\matrix{-2 & 1 \cr  1 & 0}\right)$$
which, after a change of basis, is identical to $U$. This means that the
necessary  condition for a $K3$ of Weierstra\ss\ type is that the Picard
lattice must contain $U$ as a sublattice. This also turns out to be a
sufficient condition for the $K3$ to admit an elliptic fibration.

It is clear from the discussion above that the two elements of
$H^2(K3,R)$ dual to the base and the fiber span $U$.   Thus if $G_4$
must have one leg along the base and the other along the fibre of $K3_2$
it cannot contain any element of the sublattice $U$. Since the matrix
$U$ has eigenvalues $(1,-1)$, this means $G_4$  must be orthogonal
to all vectors in $U$.

\appendix{B}{The $(-E_8) \oplus (-E_8)$ Lattice}

Here we discuss the $(-E_8) \oplus (-E_8)$ lattice in more detail. The
discussion will be based on section 11.6 in Polchinski's book
\Polchinski\ with some changes
in convention, to account for the time-like nature of the lattice in our
case etc. As mentioned in the discussion above we take  $16$ vectors,
$E_I, I=1, \cdots 16$ that are linearly independent and satisfy the  the
conditions $E_I \cdot E_J=-\delta_{IJ}$. These vectors span the $(-E_8)
\oplus (-E_8)$ vector space. The first $(-E_8)$ lattice is   given by
vectors of the form $\sum_{I=1}^{8} q^IE_I$, where
the $(q^1, \cdots q^8)$ are either all integer or half integer and
$\sum_Iq^I \in 2 {\bf Z}$. The second $(-E_8)$ lattice is similarly  given
by appropriate
 linear combinations of $E_I, I=9, \cdots 16$.

It is now easy to see why $2 \eto, 2 \ett,$ \defeto,\defett,  belong to
the lattice $\Gamma^{3,19}$. ${\tilde e}_1, n_1$ belong to the first $U$
sublattice of $\Gamma^{3,19}$, as we discussed in section 3 and 4,
and similarly ${\tilde e}_2, n_2$ belong to the second $U$ sublattice. For
the choice \wa, \wb, we have $W_IW_I={\tilde W}_I {\tilde W}_I=2$. So
to show that $2 \eto, 2 \ett,$ belong to $\Gamma^{3,19}$ it is enough to
show that $2 W_I E_I$ and $2 {\tilde W}_I E_I$ belong to it.
But we see that both of these vectors are of the form $q^IE_I$ with
$q^I$ meeting the conditions mentioned above, so this is true.

\appendix{C}{Flux Breaks the $M5$ Brane Worldvolume $U(1)$ Symmetry}
As discussed in \S6  the arithmetic genus of the divisor wrapped by
the Euclidean $M5$ brane
 corresponds to an index graded by the  $U(1)$ symmetry which is the
 structure group of the normal bundle of the
divisor.  Here we  show that in   the $K3_1\times K3_2$ case this $U(1)$  symmetry is
broken by the $G_4$ flux.

  Roughly speaking the argument is as follows. The
divisor is $K3_1 \times P^1$, with the $P^1$ being an  exceptional
divisor of $K3_2$ associated with blowing up an ADE singularity. As  we saw
above  the flux, $G_4$, must have one leg along the fiber and one along
the base of $K3_2$. Now
  the singularity is located at a particular
point in the base of $K3_2$ so  the tangent along the base is normal to
the divisor. Thus $G_4$ breaks the $U(1)$ symmetry.

More precisely, let us illustrate this by taking the case of an $A_1$ singularity.
In the vicinity of the singularity, $K3_2$ is described  by the equation,
$$y^2=x^2-z^2,$$
in $C^3$, with the  coordinate along the base of $K3_2$ being say,  $z$.
Now the resolved $K3_2$  is partially covered  by the coordinates $z, s_2,s_1,$ where
$y=s_2 z$, $x=s_1z$, and $s_2, s_1,$ satisfy the relation
\eqn\one{s_2^2=s_1^2-1.}
And the exceptional divisor in $K3_2$ is partially covered by $z=0$ with $s_2,s_1,$ satisfying \one
\foot{
These coordinates miss two points
 in the resolved $K3_2$. The divisor is
 actually the surface $s^2=s_1^2-s_3^2$ in $P^2$, and the two points not included  are $s_3=0, s_2=
\pm s_1$.}.

Away from $z=0$, the tangent vector along the the base $P^1$ is given by
$\partial_z$, and the tangent vector along the fiber $T^2$ by $\partial_{s_1} + {\partial s_2 \over \partial s_1} \partial_{s_1}$.
So it is  clear that as $z \rightarrow 0$, and we approach the  divisor,  the tangent vector along the
 fiber becames the tangent vector to the divisor  and the tangent vector along the base becames the
normal to the divisor. Since, as we mentioned above, $G_4$ has one leg along the base and one along the
fiber of $K3_2$, we see that the flux breaks the $U(1)$ symmetry.

The argument above  clearly generalises to  other ADE singularities in $K3_2$ as well.

\listrefs
\end